# Spin-Orbit Torque-Assisted Detection of the Canted Magnetization Phase in a CoTb-Based Ferrimagnet

Maksim Stebliy[1*], Zhimba Namsaraev[1], Michail Bazrov[1], Mikhail Letushev[1], Valerii Antonov[1], Aleksei Kozlov[1], Ekaterina Steblii[1], Aleksandr Davydenko[1], Alexey Ognev[1,5], Teruo Ono[2,3,4], Alexander Samardak[1,5]

[1]Laboratory of Spin-Orbitronics, Institute of High Technologies and Advanced Materials, Far Eastern Federal University, Vladivostok 690950, Russia
[2]Institute for Chemical Research, Kyoto University, Gokasho, Uji, Kyoto 611-0011, Japan
[3]Center for Spintronics Research Network, Graduate School of Engineering Science, Osaka University, Machikaneyama 1-3, Toyonaka, Osaka 560-8531, Japan
[4]Center for Spintronics Research Network, Institute for Chemical Research, Kyoto University, Gokasho, Uji, Kyoto 611-0011, Japan
[5]Sakhalin State University, Yuzhno-Sakhalinsk 693000, Russia

[*]stebliyme@gmail.com

**ABSTRACT**

Ferrimagnets have the potential to play a key role in spintronics *due to their high stability, low energy consumption, and rapid magnetic state switching. These characteristics are typically observed in ferrimagnetic materials near magnetic or angular compensation states. Near the magnetic compensation point, an external field can disrupt the collinearity between the sublattices, leading to aligned magnetic projections. In this work, a violation of antiferromagnetic ordering is detected by a change in the direction of the effective field induced by spin-orbit torque, without altering the dominance type. In the studied W/$Co_{70}Tb_{30}$/Ru structure, the canted phase region is observed near room temperature under external fields of approximately 0.1 T. Using macrospin simulations and analytical derivations, a correlation is established between anisotropy, interlattice exchange interaction, and the presence of the canted phase region.*

## 1. INTRODUCTION

Spintronic components must combine high magnetic state stability with rapid and efficient current-induced switching. Recent research shows that such components can be created using ferrimagnetic materials (FIM) [1] [2]. Experimental evidence indicates that the effective fields generated by spin–orbit torque (SOT) in these materials [3] [4] [5] [6] can exceed those in ferromagnets [7]. Moreover, switching times can reach sub-nanosecond levels [8] [1] [9], and domain wall movement speeds can approach several kilometers per second [10] [11]. These advantages are observed near the magnetic or angular momentum compensation state [12] [13].

In amorphous two-component ferrimagnets, compensation states can be achieved by modifying parameters such as temperature [4], atom ratio [14], or thickness [15] [16] [17]. Localized compensation states can also be obtained through non-uniform heating or variations in concentration [18] [19]. In addition, ferrimagnets exhibit significant properties, including bulk perpendicular anisotropy [20] [21], the Dzyaloshinskii–Moriya interaction [22] [23], and the potential for inducing internal spin–orbit torque [24] [25].

In the vicinity of the magnetic compensation state, the violation of antiferromagnetic ordering between magnetic sublattices (also known as spin-flop, oblique, canted, or non-collinear alignment) can be observed. Previous experimental [26] [27] and theoretical [28] [29] studies have investigated these states under strong magnetic fields. In this work, by fine-tuning the parameters, the canted phase was detected near room temperature (RT) at a field of approximately 0.1 T. This made it possible to study the behavior of the current-induced field during transitions through three phases: the Tb-dominated region, the noncollinear ordering region, and the Co-dominated region.

The work included the preparation of a series of films with the composition W(4)/$Co_{1-x}$Tbx(y)/Ru (2 nm). The concentration of Tb atoms (x) ranged from 20% to 40%, while the thickness (y) of the ferrimagnetic alloy varied from 2 to 8 nm. To analyze the behavior of the current-induced field, Hall bars with a width of 20 µm were fabricated and studied into a setup that enabled the simultaneous generation of out-of-plane and in-plane magnetic fields, as well as sample heating, as shown in Fig.1a. The technical details of the experiment are presented in the Materials and Methods section.

## 2. RESULTS

The results are organized into four parts. In Section 2.1, the magnetic properties of the CoTb alloy are studied as a function of Tb atom concentration to identify the composition near the magnetic compensation state. For the selected sample, $Co_{70}Tb_{30}$, the dependence of the current-induced field on the magnitude of the applied current (Section 2.2) and on the sample temperature (Section 2.3) is investigated to induce a transition through the canted phase. Section 2.4 presents the results of macrospin simulations and an analytical description of the canted phase in a ferrimagnet.

### 2.1. EFFECT OF SAMPLE COMPOSITION ON THE CURRENT-INDUCED FIELD

The saturation magnetization ($M_s$) of CoTb(2.5 nm) films was studied as a function of Tb atom concentration using a vibrating sample magnetometer (VSM) at RT, Fig.1b. The concentration range defines the region with perpendicular magnetic anisotropy. A state of magnetic compensation is observed at a Tb atomic concentration of 27%. To the left of this point, the alloy is expected to be Co-dominated: the magnetic moments of the Co sublattice align with the external magnetic field, while those of the Tb sublattice align oppositely. Conversely, on the right side of the compensation point, Tb dominance is expected. Hysteresis loops obtained from anomalous Hall voltage (AHV) measurements confirm this assumption by exhibiting a change in chirality on either side of the compensation point (inset to Fig.1b). Notably, in the ferrimagnetic material CoTb, magnetotransport, magneto-optical phenomena, and interactions with spin currents occur exclusively through the Co magnetic sublattice [30] [31]. This disparity arises because the 3d electrons, which determine the magnetic moment of Co atoms, are located closer to the Fermi level than the 4f electrons responsible for the magnetic moment of Tb atoms [32].

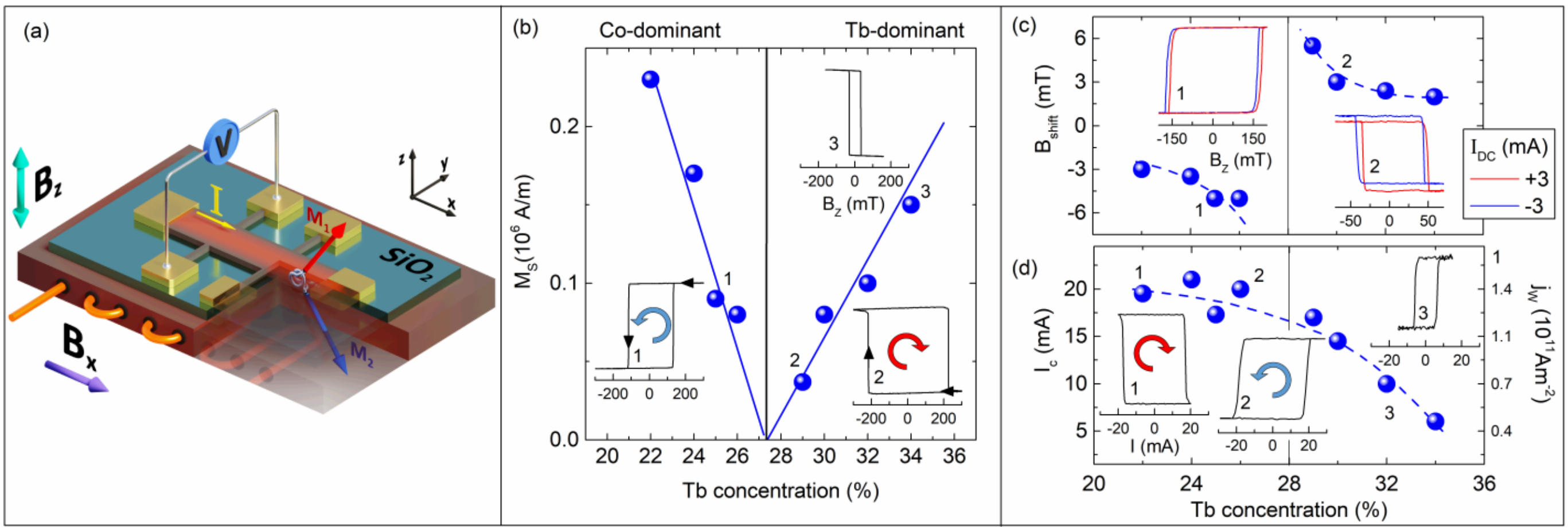

*Fig.1. (a) Schematic diagram of the experimental setup, which allows for the simultaneous passage of an electric current through the Hall bar, the application of orthogonal fields $B_x$ and $B_z$, and temperature changes using a resistive heater. (b) Dependence of saturation magnetization on the composition of CoTb (2.5 nm) films, measured using VSM. The insets show the hysteresis loops obtained by measuring the anomalous Hall effect. (c) Dependence of the perpendicular component of the current-induced magnetic field on the composition of the sample, measured using a direct current of ±3 mA in the presence of a field ($B_x$ = 0.1 T). The inset shows the loops obtained with different current orientations. (d) Dependence of the magnetization reversal current on the composition of the sample. The insets show examples of hysteresis loops observed during current-induced magnetization reversal for the highlighted cases.*

The z-component of the effective magnetic field induced by the SOT effect was measured for each sample at room temperature (RT) using the hysteresis loop shift method [33], Fig.1c. In this method, the Hall bar is remagnetized with an external field applied along the z-axis, while a direct current (±3 mA) and a fixed field along the x-axis ($B_x$ = 0.1T) are simultaneously applied. This combination generates a steady effective magnetic field along the z-axis due to the SOT effect ($\pm B_{SOT}$), resulting in a displacement of the hysteresis loop to the left or right depending on the current orientation (inset in Fig.1c).

The dependence shown in Fig.1c indicates that the direction of the current-induced field reverses upon reaching the compensation point. The alignment of the effective field is determined by the vector product of the Co-sublattice magnetization ($M_{Co}$) and the spin-current polarization (p) $\boldsymbol{B} \sim \boldsymbol{M_{Co}} \times \boldsymbol{p}$ [7]. The propagation of electric current along the +x direction generates a spin current in the W layer that is polarized along the -y direction due to the spin Hall effect. In the case of Co dominance, the external field +$B_x$ tilts the corresponding magnetic moments along +x, resulting in a current-induced field with a component along -z, as shown in the inset of Fig.2a. When the state shifts to Tb dominance, the slope of the Co magnetization changes toward -x under the same +$B_x$ field, reversing the orientation of the current-induced field to +z. Near the compensation point, an increase in the efficiency of field generation is observed (Fig.1c), consistent with previously reported results [14] [3].

Next, magnetization reversal induced by current was investigated. Current pulses with increasing amplitude and a duration of 0.1 ms were applied to the structure in the presence of a constant field of $B_x$ = +0.1 T. After each pulse, the magnetic state was evaluated using AHE signal measurements. The dependence of the critical current required to switch the magnetic orientation on Tb atom concentration is presented in Fig.1d. The current value is also converted into current density, accounting for the resistance distribution between the layers of the structure, as detailed in Section S.1 of the Supplementary Material.

The magnetic hysteresis loop from current-induced remagnetization exhibits altered chirality with a shift in dominance type [14]. In this system, chirality changes before reaching the compensation concentration, as indicated by the insets of Fig.1d. This change in dominance is likely due to Joule heating, where rising temperatures reduce the Tb sublattice moment more rapidly than that of Co [13] [34]. Consequently, increasing temperature at a fixed concentration can shift the system from Tb-dominance to Co-dominance [35] [18].

### 2.2. Effect of Current on the SOT Field

To systematically investigate the influence of current magnitude on the induced field, the loop shift study was repeated for currents ranging from 1 to 20 mA (Fig.2a). Before examining the dependence for the transient 30% concentration, we first consider the cases of 34% and 25% concentration. Within the studied range, these samples exhibited Tb and Co dominance, respectively. In both cases, an increase in the transmitted current led to a linear rise in the $B_z^{SOT}$ field, consistent with previous research [36] [14].

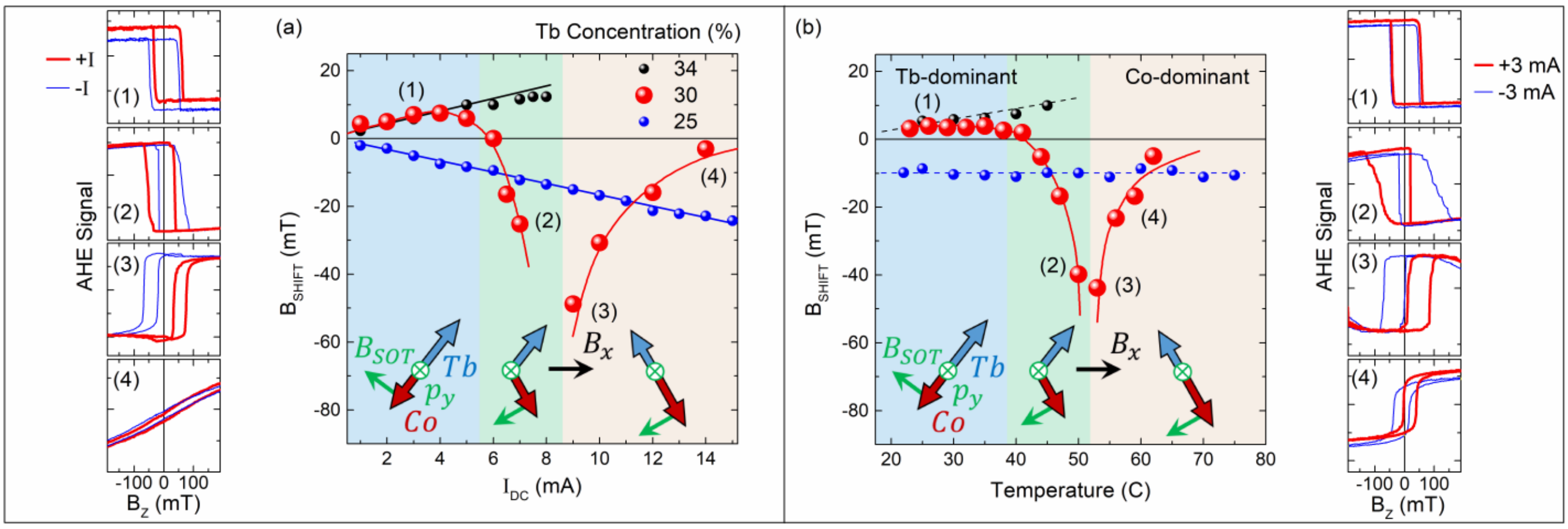

*Fig.2. (a) Dependence of the loop shifts on the magnitude of the transmitted DC current for samples with varying atomic compositions under the influence of an in-plane DC field $B_x = +0.1\ T$. The green region indicates non-collinear spin alignment in the Co and Tb sublattices. Schematics illustrate the orientation of the effective field exerted on the Co sublattice as a consequence of the SOT effect. (b)*

*Correlation between the shift value of the loops and the stage temperature at a constant current of ±3 mA and field $B_x = +0.1\ T$. Corresponding hysteresis loops for cases shown in red in diagrams (a) and (b) are shown in insets (1)–(4).*

The case of 30% Tb concentration reveals an unexpected relationship. Within the 1–4 mA range, the sample remains Tb-dominated, as indicated by the clockwise chirality of the loop. Increasing the current in this range shifts the loop toward positive fields (inset (1), Fig. 2a), suggesting the emergence of a $+B_z^{SOT}$ field. However, beyond 4 mA, the shift decreases, reaching zero at 5.5 mA. Further increasing the current reverses the shift toward negative fields, implying the induction of a $-B_z^{SOT}$ field in the opposite direction (inset (2), Fig.2a).

The constant chirality of the hysteresis loops confirms persistent Tb dominance in this current range. Thus, the reversal of the current-induced field is solely due to changes in the Co sublattice tilt under the in-plane field (green region, Fig.2a). The magnetization of the Co sublattice is weaker than that of the Tb sublattice, with their Mz projections antiparallel, as observed in the hysteresis loops, and their Mx projections parallel, as indicated by the direction of the current-induced field. Kerr microscopy analysis of cases (1) and (2) (Section S.2) reveals no qualitative differences in the magnetization reversal mechanism: reversal begins at the same point and proceeds via domain expansion.

An increase in propagating current changes the chirality of the hysteresis loops (inset (3), Fig.2a), confirming a shift in dominance to Co. However, the orientation of the induced $-B_z^{SOT}$ field remains unchanged. Notably, at 9.5 mA, the loop shift surpasses the coercive force, allowing magnetization to switch to a remanent state by reversing the current direction. This current level closely aligns with the switching current in Fig.1d, though it is slightly lower for DC current. Further increasing the current reduces the $-B_z^{SOT}$ field and perpendicular anisotropy, transforming the loop shape (inset (4), Fig.2a).

The dependence of the $-B_z^{SOT}$ field on DC current in the $Co_{70}Tb_{30}$ structure reveals key features. Beyond the Tb- and Co-dominant states, a noncollinear (canted or spin-flop) alignment of sublattice magnetization is possible. Near the magnetic compensation state, field generation efficiency asymptotically increases under noncollinear ordering. Since current induces both SOT and Joule heating, their effects were analysed by studying the $B_z^{SOT}$ field dependence on sample temperature under constant current.

### 2.3. Effect of Sample Temperature on the SOT Field

Using the loop shift method, the field magnitude was determined with a ±3 mA current in the presence of an external field $B_x = +0.1\ T$. Since the low current prevents sample heating, the sample temperature depends solely on the heating stage, Fig.1a. Figure 2b shows the field dependency for three concentrations: at 34% and 25%, the field magnitude remains constant, whereas at 30% it mirrors previous results. Between room temperature and 52°C, a clockwise hysteresis loop (right inset, states 1 and 2) indicates Tb-dominance. At 43°C, the $B_z^{SOT}$ field shifts from negative to positive, attributed to the formation of a canted phase. At 52°C—aligned with the magnetic compensation

temperature—the chirality changes (right inset, state 3), indicating Co-dominance. Near the compensation state, the field increases notably before diminishing with further temperature rise (right inset, state 4). Normalizing the field to the current allows quantitative comparison between experiments (see Section S.3 of the Supplementary Material).

In our analysis, we assumed that magnetotransport effects arise solely from the Co sublattice, although studies suggest that the rare-earth sublattice may also contribute to SOT [3] [2] and AHE [37]. This assumption does not alter our qualitative conclusions regarding the tilted phase. Comparing states (1) and (2) in Figs.2a and 2b reveals that the magnetization of the Tb sublattice remains unchanged along $Bx$, meaning its contribution to the SOT field is constant. Therefore, the observed change in the SOT field direction can only be attributed to a change in the tilt of the Co sublattice.

### 2.4. MACROSPIN SIMULATION AND ANALYTICAL ANALYSIS

A qualitative macrospin model simulation was conducted using MuMax3 [38] to investigate the formation of non-collinear magnetization alignment in a ferrimagnet. Due to the complexity of describing a ferrimagnet, where the resulting magnetization depends on atomic composition and temperature, a simplified model was used. The macrospin model included two magnetic moments ($M_1$ and $M_2$) coupled via an antiferromagnetic exchange interaction ($A_{ex}$), as shown in Fig.3a. The magnetostatic interactions and anisotropy were neglected. The simulation script is provided in Section S.7 of the Supplementary Material.

The equilibrium state of this system is governed by two interactions: with an external field ($B_x$) and exchange coupling. The Bx field generates a torque ($\tau_B^y \sim M^z B_x$) that aligns the moments parallel to the field, minimizing Zeeman energy. Conversely, exchange interaction produces a torque ($\tau_{ex}^y \sim J_{12} sin\alpha$) that favors antiparallel alignment to minimize exchange energy [38] [30].

Near the compensation point, the sublattice magnetizations are assumed to vary linearly with temperature [13] [34]. This assumption implies that $M_1$ and $M_2$ change as shown in Fig.3b, with the approximations $M_1 = (1.2 - 5 \cdot 10^{-3} \cdot i) \cdot 10^6 A/m$ and $M_2 = (1.6 - 15 \cdot 10^{-3} \cdot i) \cdot 10^6 A/m$. The temperature-dependent decrease occurs at different rates, characteristic of ferromagnetic and rare-earth sublattices. Within this range, 40 data points were analyzed, with the magnetic compensation state occurring at the midpoint. At each iteration, the system energy was minimized under a constant field $B_x = +0.125\ T$.

For an exchange interaction energy of $A_{ex} = -8 \cdot 10^{-12}\ J/m$, Fig.3 illustrates key trends: magnetic moment projections along the x-axis (c) and y-axis (d), and the angle between the moments (e). When sublattice magnetizations differ significantly, they align antiparallel along the external field (states 1 and 5, Fig.3c). As this difference decreases, the moments gradually tilt toward a perpendicular orientation to minimize Zeeman energy, creating a nonzero z-axis projection. The identical z-projections for both sublattices (Fig.3d) ensure equilibrium, as torques $\tau_{B1}$ and $\tau_{B2}$ are equal in magnitude but opposite in direction. Consequently, the projections $M_1^x$ and $M_2^x$ remain different except at the magnetic compensation point (Fig.3e), with the net magnetization along the x-axis always aligned with the field and never zero.

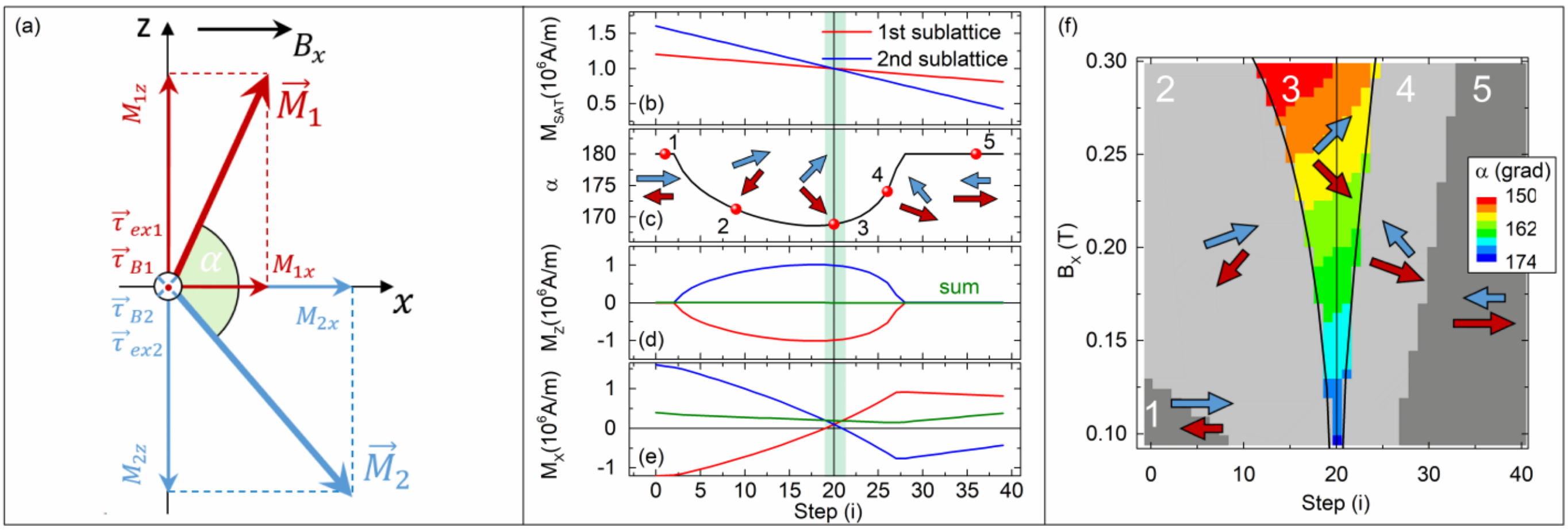

*Fig.3. (a) Schematic illustration of the macrospin model, showing the mutual alignment of magnetization in magnetic sublattices 1 and 2, along with their projections onto the x- and z-axes, and the torques induced by an external field $\tau_B$ and antiferromagnetic exchange interaction $\tau_{ex}$. To account for temperature variations, a linear reduction in the saturation magnetization of both sublattices is implemented, as shown in (b). (c) Displays the angle between the magnetizations, while (d) and (e) show the corresponding adjustments in the magnetization projections onto the x- and z-axes. The total of the magnetization projections is denoted by a green line, with the light green region representing the inclination of both magnetic moments along the field direction. (f) Shows the region of non-collinear alignment as a function of the external field magnitude. The black line indicates the field values $B_x^c$ derived from the analytical model.*

At magnetic compensation (state 3), the system is symmetric, with equal projections $M_1^z$ and $M_2^z$ and $M_1^x$ and $M_2^x$. In this balanced state, neither sublattice dominates, eliminating the need for opposition to the field. However, it does not correspond to the minimum angle between magnetic moments (Fig.3e) or energy extremes (Section S.4). Near compensation, a key region of antiferromagnetic ordering violation emerges (marked in green), where z-axis projections remain antiparallel while x-axis projections align.

Similar trends to Fig.3c were observed across various external fields and are summarized in diagram (Fig.3f). As field strength increases, the region with co-directional projections expands, aligning qualitatively with previous theoretical studies [28] [29].

A simplified ferrimagnet model enabled an analytical description of the system (Section S.5). Using Equation (1), the critical external field ($B_x^c$) values defining the boundary of antiferromagnetic ordering violation can be determined. As shown in Fig.3f and Fig.S4, the results align fully with simulations. The linear dependence of $B_x^c$ on the antiferromagnetic exchange interaction is evident at constant magnetic moments, with the co-directional x-projection region shrinking as $A_{ex}$ increases. Similar phase diagrams have been reported using the molecular-field approximation method [39] [40].

$$B_x^c = kA_{ex}\left(\left|\frac{1}{M_2^2} - \frac{1}{M_1^2}\right|\right)^{\frac{1}{2}} \qquad (1)$$

The model becomes more complex when incorporating perpendicular magnetic anisotropy in sublattices 1 and 2, leading to discrepancies in their z-axis projections (Section S.4). As anisotropy increases, the range of coaxial projection orientations expands, though the inclination angle within this range decreases (Section S.4). Analytical formulations have been derived for these scenarios: Equation (2) applies to the canted phase left of the magnetic compensation state (rare-earth dominance), while Equation (3) applies to the right (ferromagnetic dominance). The anisotropy constants for each sublattice account for the complex magnetic anisotropy in amorphous ferrimagnets.

$$B_x^c = 2A_{ex}\left(\frac{1}{M_1^2} - \left(1 + \frac{K_{U2}}{A_{ex}}\right)^2 \frac{1}{M_2^2}\right)^{1/2} \tag{2}$$

$$B_x^c = 2A_{ex}\left(\frac{1}{M_2^2} - \left(1 + \frac{K_{U1}}{A_{ex}}\right)^2 \frac{1}{M_1^2}\right)^{\frac{1}{2}} \tag{3}$$

This description aligns with simulation results (Fig.S5). Equations (2) and (3) integrate into Equation (1) under the condition $K_{U1} = K_{U2} = 0$. Prior studies have used thermodynamic potentials to produce similar phase diagrams [41].

## 3. Discussion

The violation of antiferromagnetic ordering between sublattices was experimentally detected within a narrow range of CoTb alloy thickness and concentration. Reducing the ferrimagnetic layer below 2.5 nm results in a loss of perpendicular magnetic anisotropy, while increasing thickness complicates measurements. At 4 nm, coercivity near the compensation point reaches 1 T (Section S.6), making the loop shift method impractical.

Increasing thickness appears to enhance exchange interaction energy between sublattices, as indicated by the shift of the magnetic compensation state toward lower Tb concentrations (S.6) [15] [42]. Additionally, magnetic anisotropy energy likely increases due to bulk properties of the amorphous alloy [20] [21]. Macrospin modeling suggests that higher exchange energy reduces the region of antiferromagnetic order violation, while increased anisotropy expands it.

This study identified a limited number of points in the antiferromagnetic breakdown region due to current or temperature variations. To validate the findings, measurements were repeated on new samples, yielding consistent results. Despite its limitations, this approach successfully mapped the region, whereas previous studies relied on AHE hysteresis loop analysis in fields above 3 T [26] [27] [43].

The observation of a tilted phase in a 0.1 T field near room temperature (RT) suggests practical applications, particularly enabling current-induced field reorientation in a ferrimagnet without altering dominance type. By adjusting parameters, magnetization can be manipulated by changing current magnitude while maintaining its direction and the external field. Previous instances of double switching in ferrimagnets involved shifts in dominance type [35] [44].

## 4. CONCLUSION

In this study, the relationship between the current-induced magnetic field and the sample temperature in a ferrimagnetic structure containing a CoTb alloy was experimentally investigated. In a W(4)/$Co_{70}Tb_{30}$(2.5)/Ru(2 nm) system, a phenomenon was observed in which the direction of the current-induced magnetic field could be reversed without altering the type of dominance near the magnetic moment compensation states. This behavior was attributed to the disruption of the antiferromagnetic ordering between the Co and Tb magnetic sublattices over a narrow temperature range. Furthermore, it was demonstrated that such disruptions in ordering could arise due to Joule heating, which increases with the passing current. Macrospin modeling revealed that the temperature range in which this phenomenon occurs shrinks as the interlattice exchange interaction energy increases. This finding may help explain the challenges in detecting this intermediate phase in samples with greater ferrimagnetic thickness in previous studies.

## 5. MATERIAL AND METHODS

Sample preparation: Thin films of W(4)/$Co_xTb_{1-x}$(2.5)/Ru(2), with thicknesses specified in nanometers, were prepared by magnetron sputtering at room temperature on oxidized silicon substrates Si/$SiO_2$. The sputtering process took place under an argon pressure of 0.4 Pa, with a base pressure of $1.3\times10^{-6}$ Pa. Sputtering rates for W and Ru were 0.27 Å/s (with 22 W power) and 0.49 Å/s (with 75 W power) respectively. The CoTb alloy was created through co-sputtering of Co and Tb targets inclined at a 30° angle from the sample holder's center. By adjusting the Co sputtering rate from 0.18 Å/s (at 20 W) to 0.97 Å/s (at 107 W) while maintaining a fixed sputtering rate of of $\upsilon_{Tb}$ - 0.27 Å/s (at 22 W). The percentage of atoms was calculated according to the formula $\%_{Co} = \left[1 + \frac{\upsilon_{Tb}\rho_{Co}A_{Co}}{\upsilon_{Co}\rho_{Tb}A_{Tb}}\right]^{-1} \times 100\%$, $\rho_{Co} = 8{,}9 \cdot 10^3 kg/m^3$ and $\rho_{Tb} = 8{,}22 \cdot 10^3 kg/m^3$ - the densities of materials, $A_{Co} = 58{,}9 \cdot 10^{-3} kg/mol$ and $A_{Tb} = 158{,}9 \cdot 10^{-3} kg/mol$ – the atomic weights. To ensure uniformity, samples were rotated during sputtering. Hall bar structures with 20-µm wide current guides were patterned using photolithography and lift-off techniques followed by magnetron sputtering. At the second step, the Cu(50 nm) contact pads were formed.

Magnetic Properties and Electrical Transport Characterization: The analysis of the films' magnetic characteristics was conducted using a vibrating sample magnetometer (7410 VSM, LakeShore). The resistive properties, anomalous Hall effect (AHE) magnitude, and current flow through the Hall bar were measured using a Keithley 6221 Current source and Keithley 2182A Nanovoltmeter setup. Temperature control from room temperature up to 90°C with 1°C precision was achieved through a resistive heater-equipped holder. Magnetic structure visualization was performed with a Kerr microscope (Evico Magnetics) featuring electromagnets capable of generating magnetic fields oriented both parallel and perpendicular to the sample plane.

## 6. SUPPORTING INFORMATION

Estimation of current distribution in structure layers based on experimental data; visualization of Hall bar magnetization reversal process using Kerr microscope; extended experimental data shown in Fig.2; additional results of micromagnetic simulation; analytical derivation for formulas (1-3); code for micromagnetic simulation.

## 7. ACKNOWLEDGMENTS

The work of Zh.N., M.B., M.L., A.D., A.S. was supported by Russian Science Foundation, grant No. 25-42-00083, https://rscf.ru/project/25-42-00083/. A.K. thanks Russian Science Foundation, grant No.

23-42-00076, https://rscf.ru/project/23-42-00076/. A.O., V.A. acknowledge state assignment of the Ministry of Science and Higher Education of the Russian Federation (project No. FZNS-2023-0012).

***Supporting Information***

# Spin-Orbit Torque-Assisted Detection of the Canted Magnetization Phase in a CoTb-Based Ferrimagnet

Maksim Stebliy[1*], Zhimba Namsaraev[1], Michail Bazrov[1], Mikhail Letushev[1], Valerii Antonov[1], Aleksei Kozlov[1], Ekaterina Steblii[1], Aleksandr Davydenko[1], Alexey Ognev[1,5], Teruo Ono[2,3,4], Alexander Samardak[1,5]

[1]Laboratory of Spin-Orbitronics, Institute of High Technologies and Advanced Materials, Far Eastern Federal University, Vladivostok 690950, Russia
[2]Institute for Chemical Research, Kyoto University, Gokasho, Uji, Kyoto 611-0011, Japan
[3]Center for Spintronics Research Network, Graduate School of Engineering Science, Osaka University, Machikaneyama 1-3, Toyonaka, Osaka 560-8531, Japan
[4]Center for Spintronics Research Network, Institute for Chemical Research, Kyoto University, Gokasho, Uji, Kyoto 611-0011, Japan
[5]Sakhalin State University, Yuzhno-Sakhalinsk 693000, Russia

[*]stebliyme@gmail.com

## Contents

## S.1. Estimation of current distribution in the structure

To estimate the current distribution in the layers of the W(4)/Co1-xTbx(y)/Ru(2 nm) structure, where x represents the percentage of Tb atoms and y is the thickness, the layer resistances must be known. The resistance of each layer is determined by its resistivity (ρ) and geometry, as $R = \rho l / th$, where $t$ and $h$ are the width and length of the Hall bar, respectively, Fig.S1a. The total resistance of the structure can be represented as three parallel resistances $1/R = 1/R_{Ru} + 1/R_{CoTb} + 1/R_W$, Fig.S1b. For Ru and W, resistivity ratio is 1.27 based on the tabulated values for bulk materials. The resistivity of CoTb alloy was determined experimentally by measuring the Hall barr resistance for different compositions, Fig.S1c. The resulting formula (1) was used to fit the experimental dependence of the total resistance of the structure on the thickness of layer $Co_{68}Tb_{32}$, Fig.S1c.

$$R = \frac{l}{h}\left(\frac{t_{Ru}}{1.27\rho_W} + \frac{t_{CoTb}}{\rho_{CoTb}} + \frac{t_W}{\rho_W}\right)^{-1} \quad (1)$$

As a result of the fitting process, the following resistivity values were obtained: $\rho_W = 1.1\ \mu\Omega \cdot m$, $\rho_{Ru} = 1.4\ \mu\Omega \cdot m$ and $\rho_{CoTb} = 2.2\ \mu\Omega \cdot m$. These values are an order of magnitude greater than the tabulated resistivity values for bulk materials. However, such an increase is typical for nano-sized thicknesses. Based on these values, the following layer resistances were obtained: $R_{W(4)} = 3.6\ k\Omega$, $R_{Ru(2)} = 9.1\ k\Omega$ and $R_{CoTb(2.5)} = 9.5\ k\Omega$. With a CoTb layer thickness of 2.5 nm, the following current distribution is obtained across layers: 57% of the current goes through W layer, 22% - Ru and 21% - CoTb. These obtained values were used to estimate the current density passing through the W layer, as shown in Fig.1d.

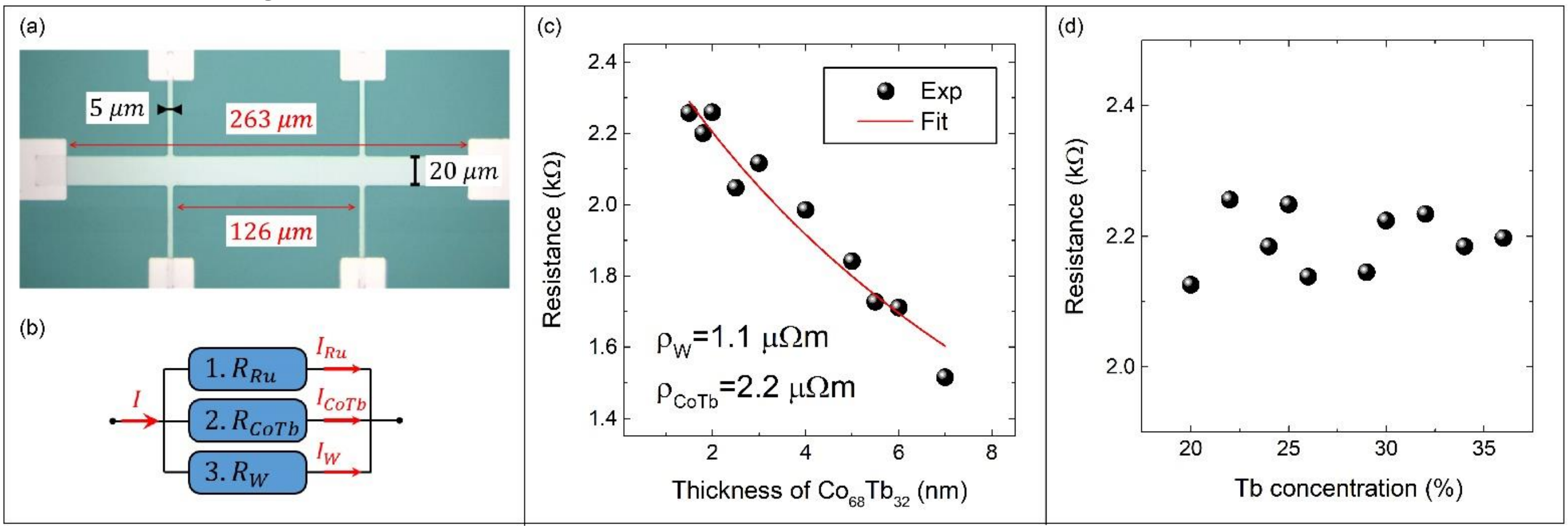

*Fig.S1. (a) Photograph of the Hall bar under study from an optical microscope with the main dimensions indicated. (b) Schematic representation of current distribution in the layers of the structure. (c) Dependence of the* W(4)/$Co_{68}Tb_{32}$(x)/Ru(2 nm) *structure resistance on the thickness of the ferrimagnetic layer. The red line shows the result of the fitting. (d) Dependence of the* W(4)/$Co_{1-x}Tb_x$(2.5)/Ru(2 nm) *structure resistance on the Tb atom concentration.*

Next, the effect of the composition of the ferromagnetic alloy of a fixed thickness on the resistance of the structure W(4)/$Co_{1-x}Tb_x$(2.5)/Ru(2 nm), was investigated. The results obtained did not demonstrate a dependence, Fig.S1d. The scatter in resistance values is about 5%. It was assumed that in the studied range of concentrations of Tb atoms, the resistivity of the CoTb layer assumed to be constant relative to the composition.

## S.2. Visualization of the magnetization reversal process using a Kerr microscope

A study of the process of magnetization reversal of CoTb Hall bars in the presence of direct current showed that, depending on the magnitude and direction of the current, the hysteresis loop can shift, Fig.1c. The Fig.S2. show a visualization of the magnetic structure of a sample with a 30 % Tb concentration, during magnetization reversal by an external perpendicular field $B_z$. In this case, a constant current ±3 and ±6.5 mA was passed through the sample in the presence of a constant field in the plane $B_x = 100\ mT$. The conditions considered approximately correspond to cases (1) and (2) in Fig.2a. As can be seen, there is a corresponding shift in the hysteresis loops when the direction of the current changes Fig.S2a-b. However, in all cases, the course of the magnetization reversal process is the same (Fig.S2c-d). Based on the results obtained, we can conclude that the shifts of the loops are not associated with a qualitative change in the magnetization reversal process. The hysteresis loops shown in the figure are constructed based on color analysis of the Kerr images.

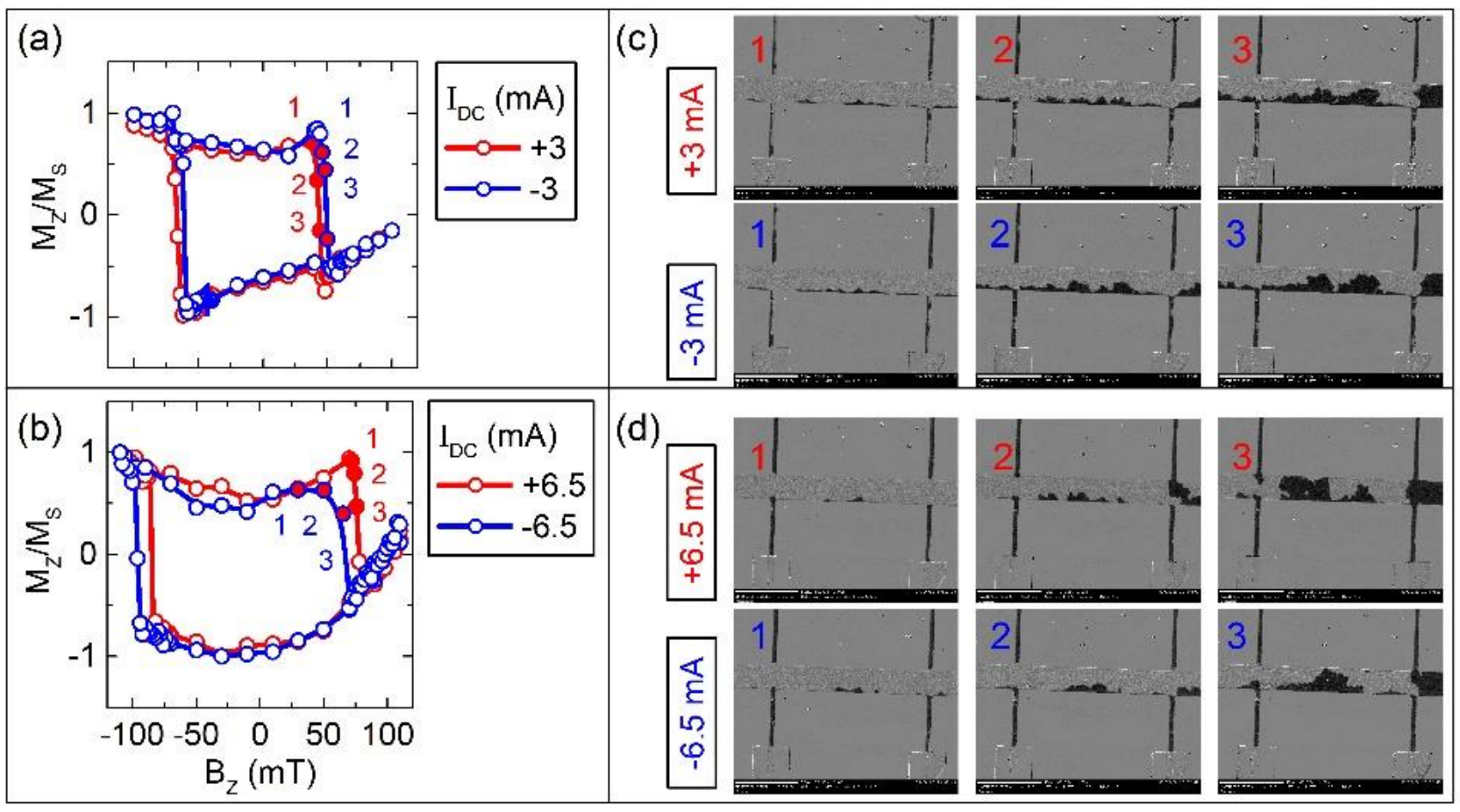

*Fig.S2. Results of a study of the magnetization reversal process of a $Co_{70}Tb_{30}$ sample during a perpendicular field sweep in the presence of a constant plane field $B_x = 100\ mT$ and direct current ±3 (a) and ±6.5 mA (b). Figures (c) and (d) show visualizations of the magnetic structure obtained by using Kerr microscopy for the marked red filling of cases.*

## S.3. The effect of the current value and temperature on the SOT field

This section contains an expanded figure compared to the figure Fig.2. in the main text. In Fig.S3, the dependence of the effective field induced as a result of the effect, normalized by the value of the transmitted current, has been added (c-d). Such a transformation is necessary to compare the results of two experiments: the influence of current (a), at a constant temperature; influence of temperature (b), at constant current. Based on the qualitative and quantitative agreement of the above results (c-d), it can be concluded that the qualitative behavior of the dependences of effective fields on the value of the transmitted current is due to a change in temperature, as a result of Joule heating. Graphs (e-f) show the dependences of the coercive force for these two experiments.

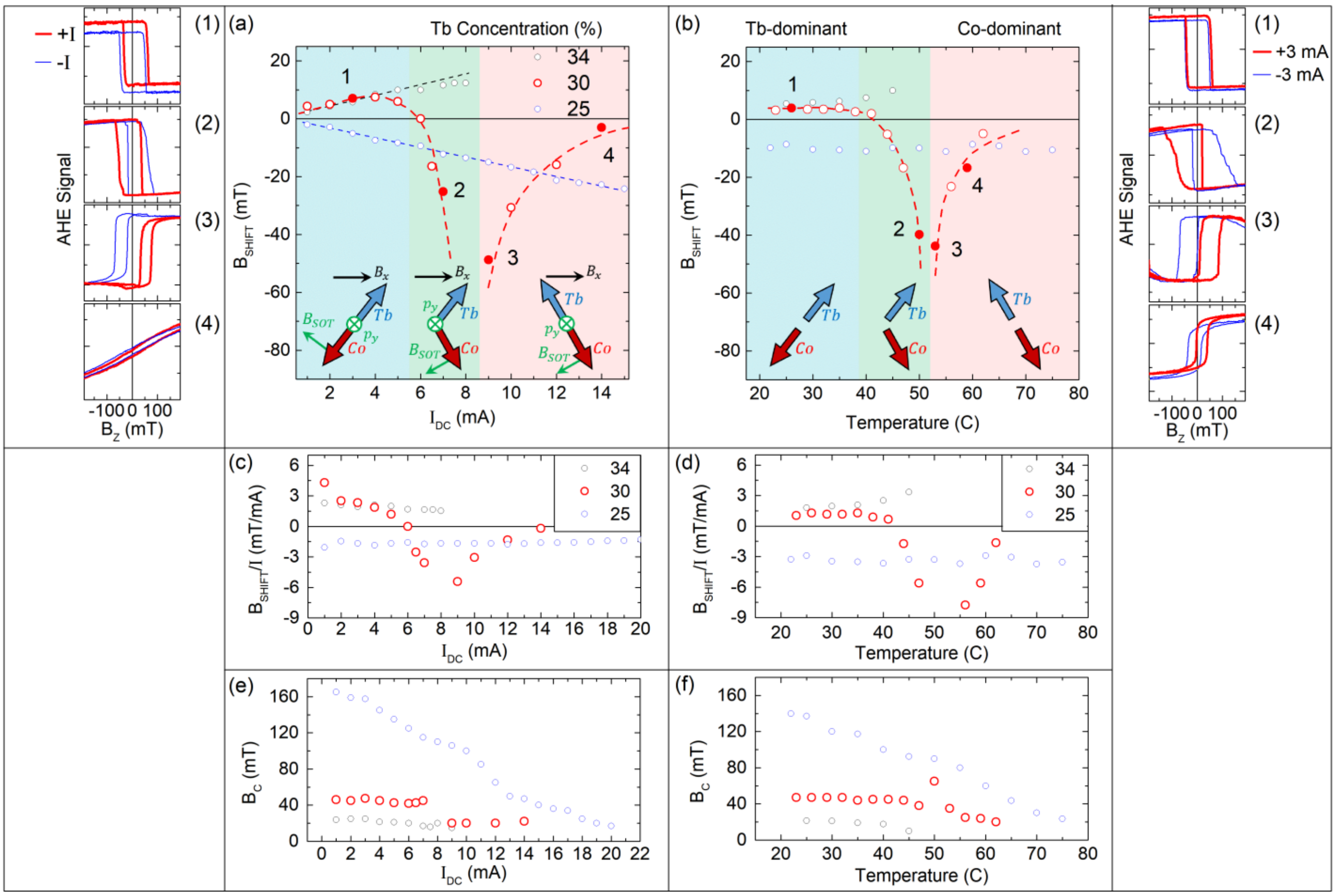

*Fig.S3. (a) Dependence of the loop shift on the magnitude of the transmitted DC current for the samples with different atomic contents obtained in the presence of the DC field in plane $B_x = 100\,mT$. The region of Tb dominance is marked in blue, the region of Co dominance in red, and the region with non-collinear alignment of spins in the Co and Tb sublattices in green. The diagrams indicate the orientation of the effective field acting on the Co sublattice in the result of SOT. Examples of hysteresis loops for cases marked in red are shown in the (1)-(4) insets. (b) Dependence of the loop shift value on the stage temperature at a fixed current ±3 mA and field $B_x = 100\,mT$. Samples with the same concentrations as in the previous case were considered. Figures (c) and (d) show the dependence of the loop shift normalized to the current value on current and temperature, respectively. Dependence of the coercive force for samples with concentrations previously considered on the magnitude of the transmitted current (e) and temperature (f).*

## S.4. Results of the macrospin simulation

This section presents extended results of the macrospin simulation described in Chapter 2.4 “Macrospin simulation and analytical analysis” of the main text. First, the case of the absence of anisotropy is considered. The Fig.S4. is supplemented, in comparison with Fig.3, with the Zeeman energy and exchange energy dependences on the step, which reflects an increase in temperature. As you can see, none of these dependencies demonstrates extrema points that make it possible to determine the states of the canted phase. Diagram (h) is given in the main text. To the left and right of it are diagrams for the cases of lower (g) and higher (i) values of the antiferromagnetic exchange interaction between the sublattices.

Below the diagrams are the dependences of the inclination angles $\alpha_1$ and $\alpha_2$ of magnetic moments $M_1$ and $M_2$ relative to x-axis at a fixed value of the external field. These dependencies illustrate the obvious fact that the region of the canted phase is limited on the left and right by the consistent transition of magnetic moments $M_1$ and $M_2$ through a state with a vertical orientation $\alpha_1 = 90°$ and $\alpha_2 = 90°$. This condition was then used to obtain an analytical description of the region of existence of the canted phase.

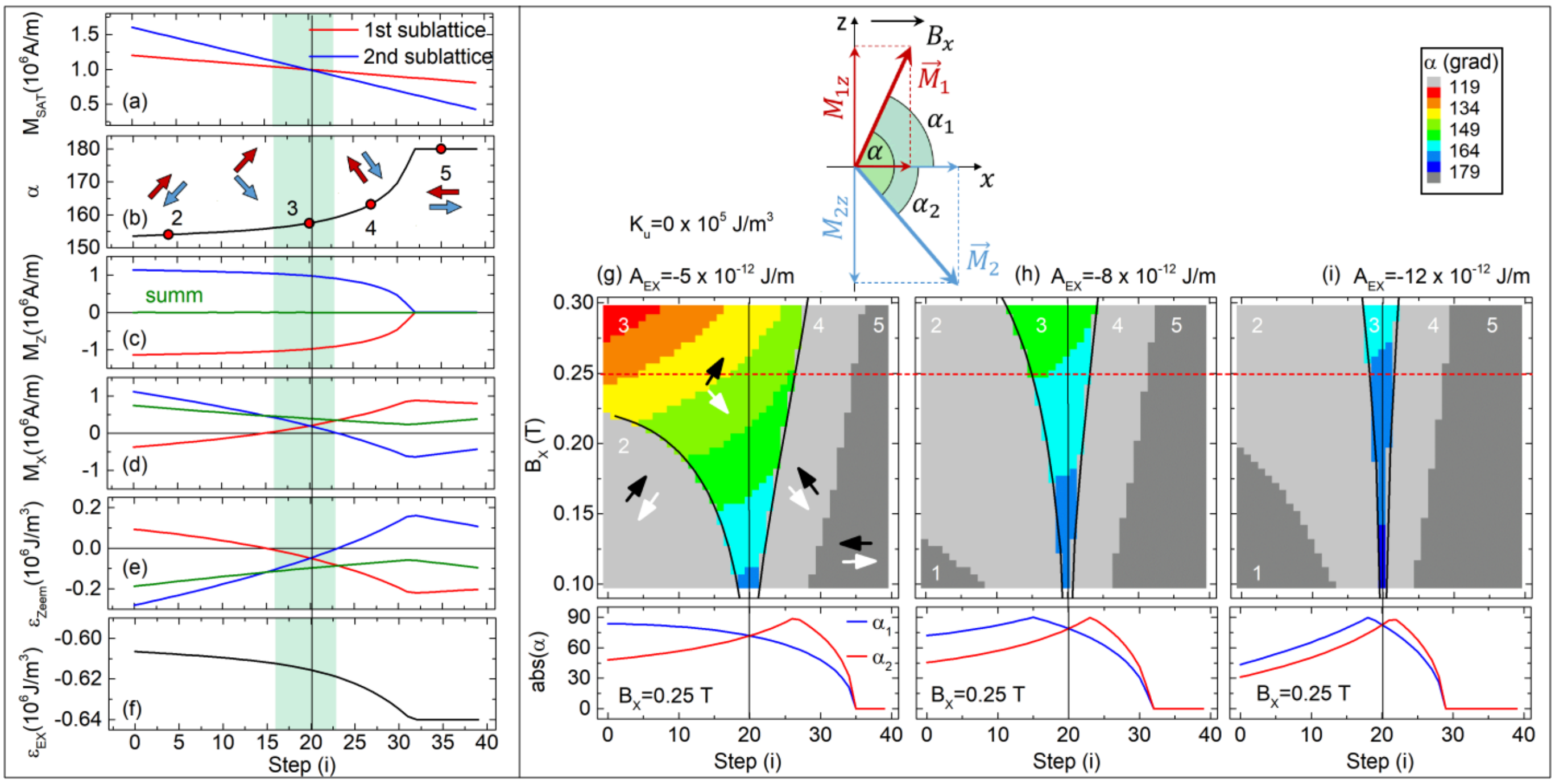

*Fig.S4. (a) The specified dependence of the magnetic moment of each of the sublattices on the step number i, reflecting the increase in temperature. (b) Dependence of the angle between the magnetic moments marked in the diagram, depending on the step, obtained during relaxation in the presence of the field $B_x = +0.125\ T$. The exchange energy is -8e-12 J/m, the anisotropy energy is zero. Plots (c-d) show the corresponding change in the projections of magnetizations onto the z and x. The sum of the magnetization projections is marked with a green line. The light green area corresponds to the slope of both magnetic moments along the field. Plots (e-f) show the corresponding dependences of the Zeeman energy and exchange interaction. The diagrams (g-i) show the area of non-collinear, canted alignment depending on the external field value for different values of exchange interaction $A_{ex}$ between the sublattices. The black line marks the field values obtained using the theoretical model. Below the diagrams are the dependences of the magnetization tilt angles $\alpha_1$ and $\alpha_2$ for field strength $B_x = +0.25\ T$, marked on the diagram with a red dotted line.*

Next, a more complicated case was considered when the system has anisotropy with the axis along the z-axis. Fig.S5a-f show the dependences of the angle between moments $M_1$ and $M_2$, their projections on the axes z and x obtained during relaxation in the presence of the $B_x = +0.125\ T$. With parameters $K_u = 0.2e5\ J/m^3$ and $A_{ex} = -8e-12\ J/m$. The previously considered diagram (g) is shown in comparison with cases of sequential increase in anisotropy (h) and (i). The region of existence of the canted phase increases. However, as can be seen in the dependences under the diagrams, with increasing anisotropy the deviation from values $\alpha_1 = 90°$ and $\alpha_2 = 90°$ decreases. That is, an increase in anisotropy is expected to lead to a vertical alignment of magnetic moments.

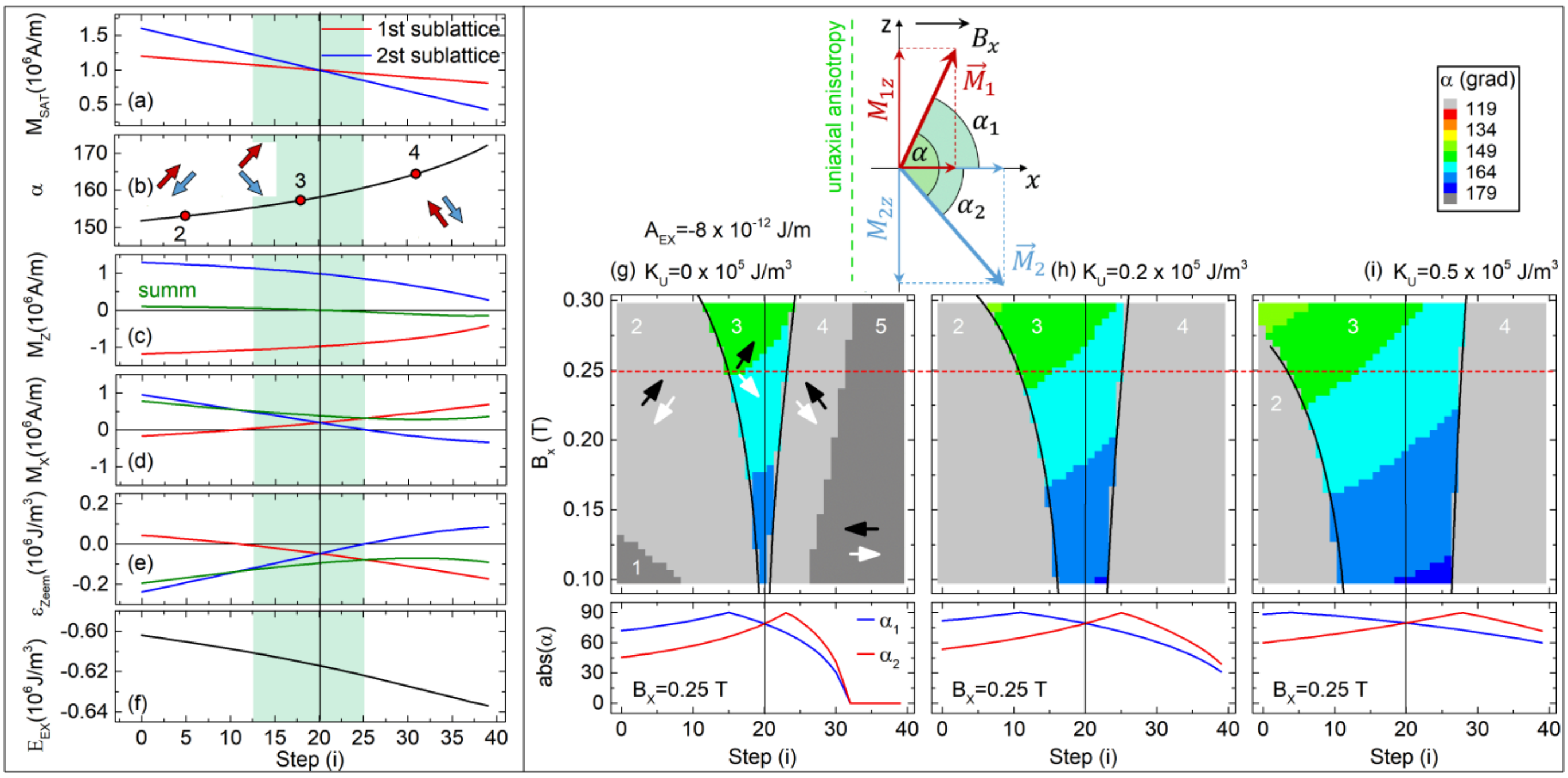

*Fig.S5. (a) The specified dependence of the magnetic moment of each of the sublattices on step number i, reflecting the increase in temperature. (b) Dependence of the angle between the magnetic moments marked in the diagram, depending on the step, obtained during relaxation in the presence of the field $B_x = +0.125\ T$. The exchange energy is -8e-12 J/m, the anisotropy energy is 0.2e5 J/m$^3$. Plots (c-d) show the corresponding change in the projections of magnetizations onto the z and x axes. The sum of the magnetization projections is marked with a green line. The light green area corresponds to the slope of both magnetic moments along the field. Plots (e-f) show the corresponding dependences of the Zeeman energy and exchange interaction. Diagrams (g-i) show the area of non-coliniar, canted alignment depending on the external field value for cases of different anisotropy energies. The anisotropy axis coincides with the z-axis. The black line marks the field values obtained using the theoretical model. Below the diagrams are the dependences of the magnetization tilt angles $\alpha_1$ and $\alpha_2$ for field strength $B_x = +0.25\ T$, marked on the diagram with a red dotted line.*

## S.5 Analytical description of a ferrimagnet in the macrospin approximation

To interpret the results of the macrospin simulation, an analytical description of the system was proposed. As in the case of simulation, two magnetic moments $M_1$ and $M_2$ were considered, Fig.S6, representing the sublattice of rare earth and ferromagnetic atoms, respectively. Magnetic moments participate in three interactions: 1) among themselves, as a result of antiferromagnetic exchange interaction; 2) with an external magnetic field; 3) with an internal effective field due to the presence of uniaxial magnetic anisotropy oriented along the z-axis.

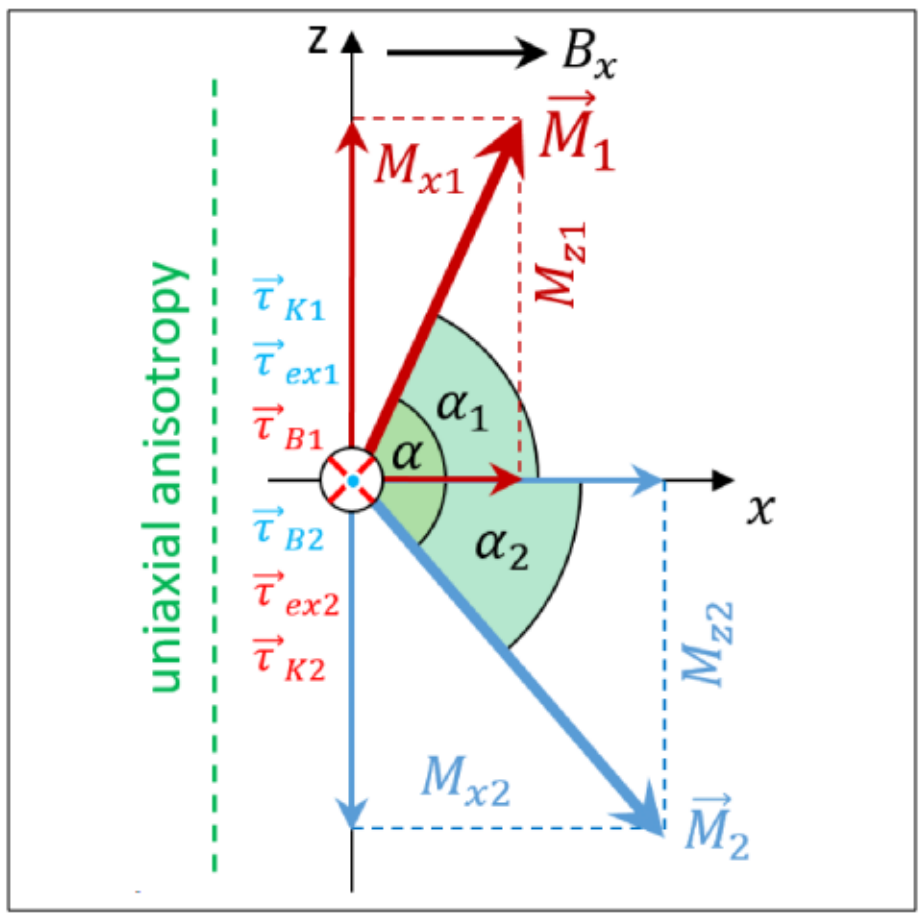

*Fig.S6. Schematic representation of the macrospin model of a ferrimagnet used for theoretical analysis. The diagram shows the mutual orientation of the magnetic moments of the rare-earth $M_1$ and ferromagnetic $M_2$ sublattices, the external field, the anisotropy axis, and the direction of the torques.*

Each of these interactions can be described through the corresponding magnetic field: 1) $\vec{B}_{ext}$, 2) $\vec{B}_{exch} = 2\frac{A_{ex}}{M_s}\Delta\vec{m}$, 3) $\vec{B}_K = 2\frac{K_U}{M_s}\vec{e}_K$. Where Aex is the exchange interaction constant between the sublattices, Ms is the magnetization of the sublattice, $\Delta\vec{m}$ is a unit vector indicating the direction of the vector of the magnetization difference in the sublattices, KU - uniaxial magnetic anisotropy constant, $\vec{e}_K$ is a vector coaxial to the anisotropy axis, and oriented in direction +z for magnetic moment M1 and in direction -z for vector M2. Taking into account the complex nature of magnetic anisotropy in amorphous ferrimagnets, for the general case the anisotropy constants are written for each sublattice. Each of the fields creates a torque acting on the magnetization, which is described by the vector product $\vec{\tau} = \vec{M} \times \vec{B}$. The equilibrium condition for the system is condition $\sum \vec{\tau}_i = 0$, written for each magnetic moment. Equations (1-3) describe the torques created by each of the described fields. Where the value of magnetic moments $M_{1(2)}$ replaces the saturation magnetization for the sublattices, and angles $\alpha_1$ and $\alpha_2$ are marked on the Fig.S6.

$$\vec{\tau}_{B1(2)} = \vec{M}_{1(2)} \times \vec{B}_{ext} \qquad (1)$$
$$\vec{\tau}_{exch} = 2A_{ex}sin(\alpha_1 + \alpha_2)\vec{e}_y \qquad (2)$$
$$\vec{\tau}_{K1(2)} = 2K_{U1(2)}sin\left(\frac{\pi}{2} - \alpha_{1(2)}\right)\vec{e}_y = 2K_{U1(2)}cos\alpha_{1(2)}\vec{e}_y \qquad (3)$$

In all cases, the torque vector is oriented along the y-axis, and equations (1-3) can be rewritten in scalar form. Using the following trigonometric relations: 1) $sin\alpha_{1(2)} = M_{z1(2)}/M_{1(2)}$, 2) $cos\alpha_{1(2)} = M_{x1(2)}/M_{1(2)}$, 3) $sin(\alpha_1 + \alpha_2) = sin\alpha_1 cos\alpha_2 + sin\alpha_2 cos\alpha_1$, trigonometric functions can be eliminated. The result will be Equations (4-6), in which the projections of magnetizations on axes x and y appear.

$$\tau_{B1(2)} = B_x M_{1(2)} sin\alpha_{1(2)} = B_x M_{z1(2)} \qquad (4)$$
$$\tau_{exch} = 2A_{ex}sin(\alpha_1 + \alpha_2) = 2A_{ex}\left(\frac{M_{z1}}{M_1}\frac{M_{x2}}{M_2} + \frac{M_{z2}}{M_2}\frac{M_{x1}}{M_1}\right) \qquad (5)$$
$$\tau_K = 2K_{U1(2)}cos\alpha_{1(2)} = 2K_{U1(2)}2\frac{M_{x1(2)}}{M_{1(2)}} \qquad (6)$$

In the case of M1, the external field tends to orient along direction +x, creating a torque along +y. The presence of anisotropy implies orientation along +z, which creates a torque along the direction -y. The exchange interaction tends to align magnetic moments antiparallel, which in the cases under consideration creates a torque along -y. In the case of $M_2$, the direction of the torques changes to the opposite. Substituting equations (4-6) into the equilibrium condition $\vec{\tau}_{B1(2)} + \vec{\tau}_{exch} + \vec{\tau}_{K1(2)} = 0$, allows to obtain the system of equations (7) describing the equilibrium state of the system.

$$\begin{cases} B_x M_{z1} - 2A_{ex}\left(\frac{M_{z1}}{M_1}\frac{M_{x2}}{M_2} + \frac{M_{z2}}{M_2}\frac{M_{x1}}{M_1}\right) - 2K_{U1}\frac{M_{x1}}{M_1} = 0 \\ B_x M_{z2} - 2A_{ex}\left(\frac{M_{z1}}{M_1}\frac{M_{x2}}{M_2} + \frac{M_{z2}}{M_2}\frac{M_{x1}}{M_1}\right) - 2K_{U2}\frac{M_{x2}}{M_2} = 0 \end{cases} \quad (7)$$

For the current study, the greatest interest is in determining the relationship between the parameters of the system, which makes it possible to obtain a state of the system with projections $M_{x1}$ and $M_{x2}$, oriented along the external field in direction +x simultaneously, which corresponds to the state of the canted phase of the ferrimagnet. To determine this area, the condition can be formulated as follows. First, it is necessary to divide the area where the sign of the projection $M_{x1}$ changes from negative to positive, which corresponds to the condition $\alpha_1 = \pi/2$ and therefore $M_{z1} = M_1, M_{x1} = 0$. Taking this into account, system (7) can be rewritten as follows.

$$\begin{cases} B_x M_1 - 2A_{ex}\frac{M_{x2}}{M_2} = 0 \\ B_x M_{z2} - 2A_{ex}\frac{M_{x2}}{M_2} - 2K_{U2}\frac{M_{x2}}{M_2} = 0 \end{cases}; \quad \begin{cases} M_{x2} = \frac{B_x}{2A_{ex}} M_1 M_2 \\ M_{z2} = \left(1 + \frac{K_{U2}}{A_{ex}}\right) M_1 \end{cases} \quad (8)$$

Using the relationship between the magnetization components $M_2^2 = M_{x2}^2 + M_{z2}^2$, system (8) can be reduced to a single equation and solved with respect to the external field. Equation (9) determines the value of the external magnetic field at which the projection $M_{x2}$ changes sign, depending on the magnetic parameters of the system. In the same way, the expression for projection $M_{x1}$, equation(10), can be obtained.

$$B_x = 2A_{ex}\left(\frac{1}{M_1^2} - \left(1 + \frac{K_{U2}}{A_{ex}}\right)^2 \frac{1}{M_2^2}\right)^{1/2} \quad (9)$$

$$B_x = 2A_{ex}\left(\frac{1}{M_2^2} - \left(1 + \frac{K_{U1}}{A_{ex}}\right)^2 \frac{1}{M_1^2}\right)^{\frac{1}{2}} \quad (10)$$

Equations (9) and (10) limit the canted phase on the left and right, respectively, in the graphs Fig.S4.2. As can be seen, in all cases, the result of the analytical assessment is in perfect agreement with the simulation results. It should be noted that in the simulation, the same anisotropy value was used for both sublattices. If there is no anisotropy in the system $K_{U1} = K_{U2} = 0$, then the pair of equations can be reduced to one using the absolute value of the difference (11). In this case, the results of the analytical description coincide with the results of the simulation in Fig.S4.1.

$$B_x = 2A_{ex}\left(\left|\frac{1}{M_2^2} - \frac{1}{M_1^2}\right|\right)^{1/2} \quad (11)$$

## S.6. Effect of CoTb layer thickness on magnetic properties

The experimental part of the work was carried out on the basis of samples with a ferrimagnetic CoTb layer thickness of 2.5 nm. However, as part of the work performed, a series of samples with different percentages of atoms were also prepared for film thicknesses of 4 and 8 nm. Figs.S7a-b show the VSM obtained dependences of the saturation magnetization and coercive force on the concentration of Tb atoms for films. As can be seen, the state of magnetic compensation shifts towards decreasing Tb content with increasing alloy thickness. This leads to the fact that at the same Tb concentration, the alloy may change from the Tb-dominante to Co-dominante state with increasing film thickness, as marked by red dots in the plot (a). As confirmation, the corresponding hysteresis loops obtained at room temperature by measuring the anomalous Hall effect are shown in Fig.S6c. This behavior can be considered as indirect confirmation that an increase in the thickness of the alloy leads to an increase in the energy of exchange interaction between the Co and Tb sublattices.

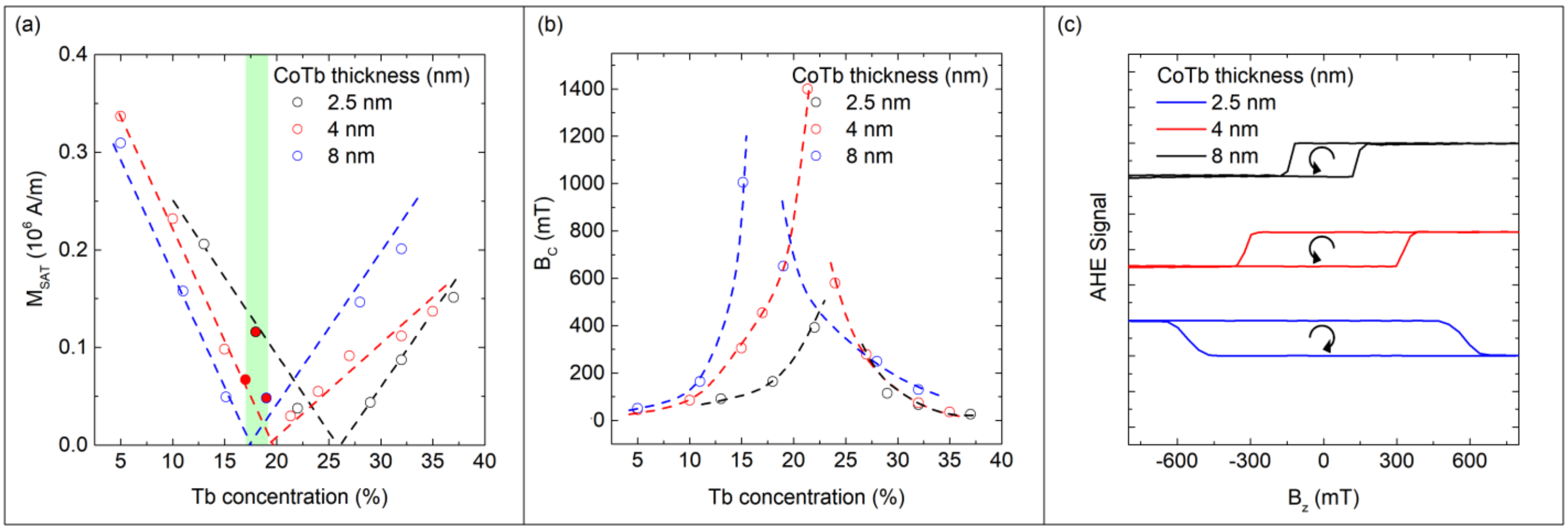

*Fig.S7. Dependence of the saturation magnetization (a) and coercive force (b) of CoTb films on the Tb atomic content at different thicknesses. The area marked in green contains cases where a change in thickness leads to a change in the type of dominance at a fixed Tb atom content. Figure (c) shows examples of loops for the corresponding cases.*

## S.7. Script used for macrospin simulation

| MuMax3 code | Comments |
|---|---|
| SetGridsize(1, 1, 2) | //Two cells are defined along the z-axis |
| SetCellsize(5e-9, 5e-9, 5e-9) | |
| i:=0 | //Step indices I and k are determined |
| k:=0 | |
| tableaddvar(i,"i","") | //Various system parameters are entered into the table |
| tableaddvar(k,"k","") | |
| tableAdd(msat.Region(1)) | |
| tableAdd(msat.Region(2)) | |
| tableAdd(m_full.Region(1)) | |
| tableAdd(m_full.Region(2)) | |
| tableAdd(B_ext) | |
| defregion(1, layer(0)) | //Each cell is defined as a separate region |
| defregion(2, layer(1)) | |
| for k=0; k<40; k=k+1{ | //The loop on parameter k iterates through the values of the external field |
| for i=0; i<40; i=i+1{ | //The loop on parameter i iterates through the saturation magnetization values |
| Mag1:=1+(20-i)*0.01 | //Dependence of layer saturation magnetization on parameter i |
| Mag2:=1+(20-i)*0.03 | |
| EnableDemag=false | //Disabling magnetostatic interaction |
| NoDemagSpins=1 | |
| Msat.setregion(1, Mag1*1e6) | //Determination of magnetic parameters of layers |
| Msat.setregion(2, Mag2*1e6) | |
| Ku1.setregion(1, 0.05e6) | |
| anisU.setRegion(1, vector(0, 0, 1)) | |
| Ku1.setregion(2, 0.05e6) | |
| anisU.setRegion(2, vector(0, 0, 1)) | |
| ext_InterExchange(1, 2, -8e-12) | //Determination of exchange interaction between cells |
| m.setRegion(1, uniform(0,0,-1)) | |
| m.setRegion(2, uniform(0,0,1)) | |
| B_ext = vector(0.1+0.005*k, 0, 0) | |
| relax() | |
| TableSave()}} | |